\documentclass[aip,pop,psfig,twocolumn,showpacs,superscriptaddress,nobibnotes,
reprint]{revtex4-1}
\usepackage{color}
\usepackage{graphics}
\usepackage{epsfig}
\usepackage[normalem]{ulem}
\usepackage{cancel}
\usepackage{float}

\usepackage[outercaption]{sidecap}

\begin{document}
\title{Observation of the Korteweg-de Vries soliton in molecular dynamics simulations of a dusty plasma medium}
\author{Sandeep Kumar}
\affiliation{Institute for Plasma Research, HBNI, Bhat, Gandhinagar - 382428, India}
\author{Sanat Kumar Tiwari}
\affiliation{Department of Physics and Astronomy, University of Iowa, Iowa city, IA, 52242, USA}
\author{Amita Das}
\affiliation{Institute for Plasma Research, HBNI, Bhat, Gandhinagar - 382428, India}
\date{\today}
\begin{abstract} 
\paragraph*{}
Excitation and the evolution of Korteweg-de Vries (KdV) solitons in a dusty plasma medium is studied using Molecular Dynamics (MD) simulation. The dusty plasma medium is modelled as a collection of dust particles interacting through Yukawa potential which takes account of dust charge screening due to the lighter electron and ion species. The collective response of such screened dust particles to an applied electric field impulse is studied here. An excitation of a perturbed positive density pulse propagating in one 
direction along with a train of negative perturbed rarefactive density oscillations in the opposite direction is observed. These observations are in accordance with evolution governed by the KdV equation. Detailed studies of (a) amplitude vs. width variation of the observed  pulse, (b) the emergence of intact separate pulses with an associated phase shift after collisional interaction amidst them etc., conclusively qualify the positive pulses observed in the simulations as KdV solitons. 
It is also observed that by increasing the strength of the electric field impulse, multiple solitonic structures get excited. The excitations of the multiple solitons are similar to the experimental observations reported recently by Boruah et al. [Phys. of Plasmas 23, 093704 (2016)] for dusty plasmas. The role of coupling parameter has also been investigated here which shows that with increasing coupling parameter, the amplitude of the solitonic pulse increases whereas its width decreases.

\end{abstract} 
\maketitle 
\section{Introduction}
\label{intro}
Solitons are robust and stable localized nonlinear structures observed in variety of natural and laboratory scenario including optical fibers\cite{Gedalin, Haus}, semiconductors\cite{Barland2002}, oceanography\cite{NEW1990513}, plasmas\cite{Zabusky, Tran, Shukla2012, Malik}, laser plasma interaction \cite{Sita, Poornakala, Kaw} etc.\cite{Sinkala, Heeger, Popel}. Mathematically, solitons are solution of non-linear equations such as Korteweg-de Vries (KdV) equation, Klein-Gordan equation and Schrodinger equation etc.\cite{Drazin, Zhang2016}. In plasmas both electrostatic\cite{Sanat_njp, Ma, Shukla_2003, Heidemann_dark, Baluku, Xue} and electromagnetic solitary waves\cite{Farina, Kaw} are observed. The robust and stable existence of solitons can be utilized for communication as well as transport of 
energy\cite{blazquez2002}. Observing solitons in ordinary electron-ion plasmas in general would require sophisticated diagnostics. However, experimental observations of solitonic structures in the context of dusty plasma can be carried out with relative ease. This is because the  temporal and spatial length scale of excitations typically lie within the perceptible grasp of human senses\cite{das2014}. 

The dusty plasma contains highly charged (mostly negative) and heavy ($10^{13} - 10^{14}$ times heavier than the ions) dust grains along with electron and ion species. The inclusion of heavy dust species makes such a plasma exhibit a rich class of collective phenomena occurring at longer time scales. There have been several experimental studies showing excitation of dust acoustic soliton \cite{Pintu, Sharma, Sheridan, Samsonov_Soliton}, their collisional interaction and the associated phase shift\cite{Sharma, Harvey}. More recently the excitation of multiple solitons have also been reported by Boruah et al.\cite{Boruah}. 
In this work we show that all these aspects can be very well depicted by treating the dust species in the plasma as particles interacting via Yukawa potential which mimics the screening due to the electron and ions species and has the following form\cite{yukawapotexp2000}:
\begin{equation} 
{U(r) = \frac{Q^2}{4 \pi \epsilon_0 r } \exp 
(-\frac{r}{\lambda_ D})}
\end{equation} 
Here $Q = -Z_d e$ is the charge on a typical dust particle, $r$ is the separation between two dust particles, $\lambda_ D$ is the Debye length of background plasma. Typical one component plasma (OCP) is characterized by two dimensionless parameters $\Gamma = \frac{Q^2}{4 \pi\epsilon_0 a k_b T_d}$ and $\kappa = \frac{a}{\lambda_D}$. Here $T_d$ and $a$ are the dust temperature and the Wigner-Seitz radius respectively. 
\paragraph*{•}
The present simulation studies employ the electric field perturbations of the experimental situations \cite{Pintu, Sharma} to excite the solitonic structures. We study the effect of the amplitude of the electric field and the width corresponding to the region where it is applied on the characteristics of the excited coherent structure. It has been observed that increasing the amplitude of the electric field paves the way for the formation of multiple solitons in the medium as observed recently by Boruah et al.\cite{Boruah}. It should be noted that by considering the response of the dust particles to the imposed electric field one makes a specific choice of the sign of the dust charge. The response of the dust species (with specified sign of the charge) to the applied electric field breaks the left and right symmetry, as a result of which one observes a positive train of solitons in one direction, whereas in the other direction rarefactive density oscillations are observed as per the KdV prescription. Our results are also in line with the experimental observations of KdV solitons by Sheridan et al.\cite{Sheridan} where authors reported a stable solitary pulse in leading direction and a dispersive wave moving in the backward direction. This is in contrast to an earlier simulation study by Tiwari et al. \cite{sanatsoliton}, where arbitrary Gaussian density perturbation splits in two solitons moving in opposite directions. This difference in observation can be understood as in the case of Tiwari et al.\cite{sanatsoliton} there was nothing in the excitation to break the left and right symmetry.

The manuscript is organized as follows. Section \ref{mdsim} provides details of simulation.  
Section \ref{solex} explains the excitation of solitonic structures and reports on various features  which confirm them as KdV solitons. In section \ref{solcol}, we discuss the collisional interaction amidst solitons. Section \ref{multisol} shows the details of the excitation of multi-solitons and effect of the coupling parameter. Section \ref{result} contains the summary.

\section{Description of MD simulations}
\label{mdsim}
Molecular Dynamics (MD) simulations have been carried out for a two-dimensional system of point dust particles interacting with each other through the Yukawa form of interaction potential. An open source classical MD code LAMMPS \cite{Plimpton19951} has been used for the purpose. A two dimensional box (with periodic boundaries) is created with $L_x = 20a$ and $L_y=1000a$ along X and Y directions respectively. Here $ a=(\pi n_{2D})^{-\frac{1}{2}} $ and  $n_{2D}$ is the dust density\cite{ZDonkoad} in two dimensions. Parameters \cite{Nosenko} chosen for the present set of simulations are as follows: the dust grain mass $m = 6.99\times10^{-13} $ Kg, charge on dust $Q = 11940e$ ($e$ is an electronic charge) and $a = 0.418\times 10^{-3}$ m. Shielding parameter $\kappa = \frac{a}{\lambda_D}$ is chosen to be 0.5 for all simulations leading to plasma Debye length to be $\lambda_D = 8.36 \times10^{-4}$ m. 
For these parameters, $E_0 = \frac{Q}{4\pi\epsilon_0a^2} = 98.39$ $\frac{V}{m}$ and equilibrium density ($n_{d0}$) $= 1.821\times 10^6$ $m^{-2}$. The cut-off for particle interaction potential in the  simulation here has been   chosen to be at $20a$. Characteristic dust plasma frequency of the particles $\omega_{pd} = \sqrt{\frac{Q^2}{2 \pi \epsilon_0 m a^3}} \simeq 35.84$ $s^{-1} $, which corresponds to the dust plasma period to be $0.175$ s. We have chosen simulation time step as $0.0072\omega_{pd}^{-1}$ s so that phenomena occurring at dust plasma frequency can be easily resolved. In this paper, distance, density, time and electric field are normalized by $a$, $n_{d0}$, $\omega_{pd}^{-1}$ and $E_0$ respectively. 

The first task is to prepare an equilibrated system. For this, the initial configuration of particle positions is chosen to be random and velocities were chosen to follow Gaussian distribution corresponding to the temperature $T_d$. Furthermore, we achieved equilibrium temperature by generating positions and velocities from canonical (NVT) ensemble using Nose-Hoover\cite{Nose, Hoover} thermostat. To test the equilibration of system, we checked temperature fluctuation and velocity distribution at different leading times. After about an NVT run for $2867\omega_{pd}^{-1}$ time, we disconnected the canonical thermostat and ran a simulation for microcanonical (NVE) ensemble for about $1433\omega_{pd}^{-1}$ time. After NVE run temperature becomes steady and equal to $T_d$. Now system is in equilibrium and ready for further explorations. For most of our studies we have chosen the value of $\Gamma = 100$ and $\kappa = 0.5$. We have, however, also studied cases with different choice of $\Gamma$.
\section{Excitation of solitons and dispersive waves} 
\label{solex}
\begin{figure}[!hbt]  
 \includegraphics[height = 7.0cm,width = 8.0cm]{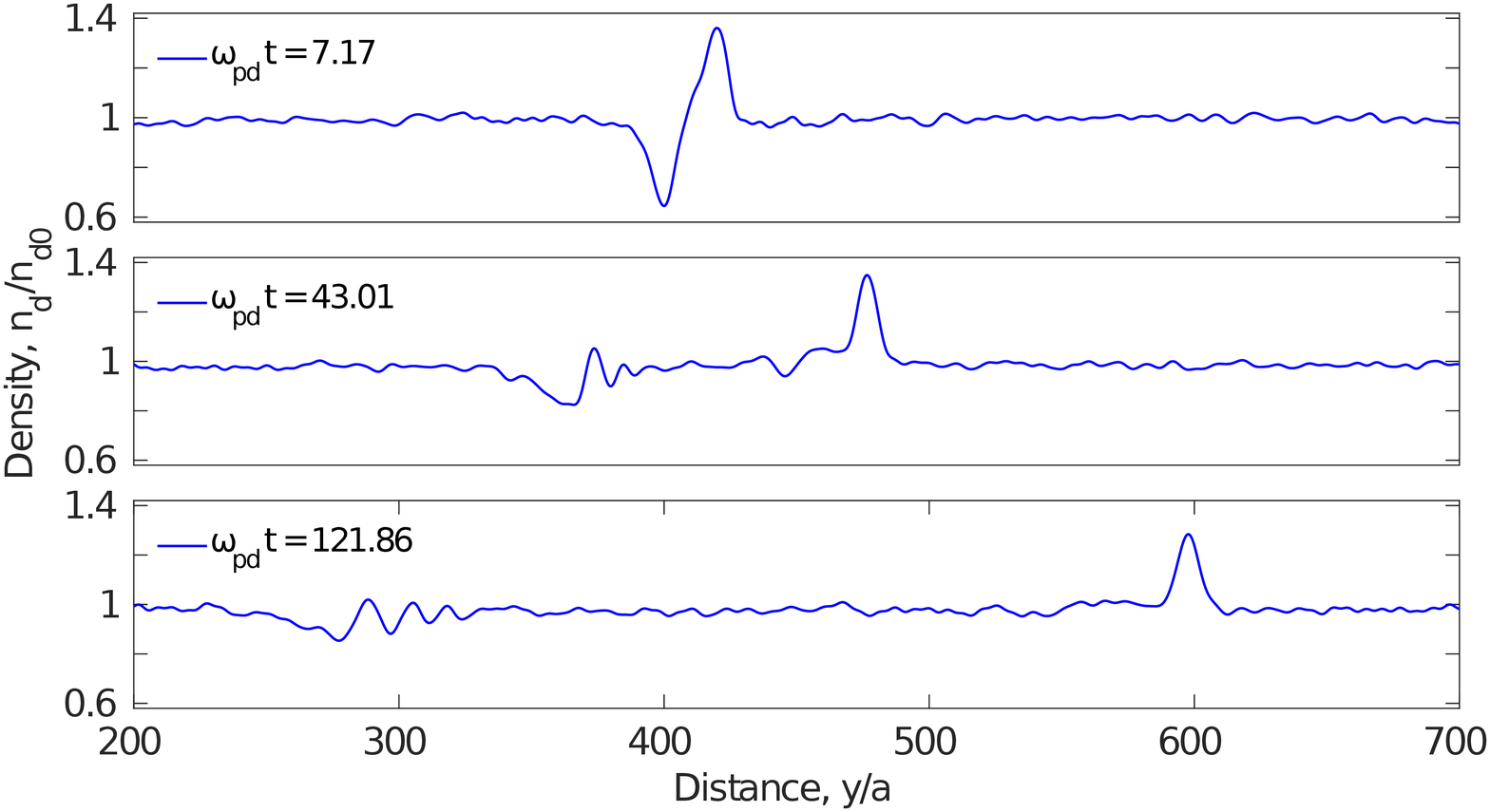}
                   \caption{Time evolution of a solitary pulse (moving  
$+\hat{y}$ direction) and a dispersive mode (moving $-\hat{y}$ direction) excited through the 
electric field perturbation ($E = 25.40$) in the medium.}
                  \label{dispersive}	                        
\end{figure}

\begin{figure}[!hbt] 
 \includegraphics[height = 7.0cm,width = 8.0cm]{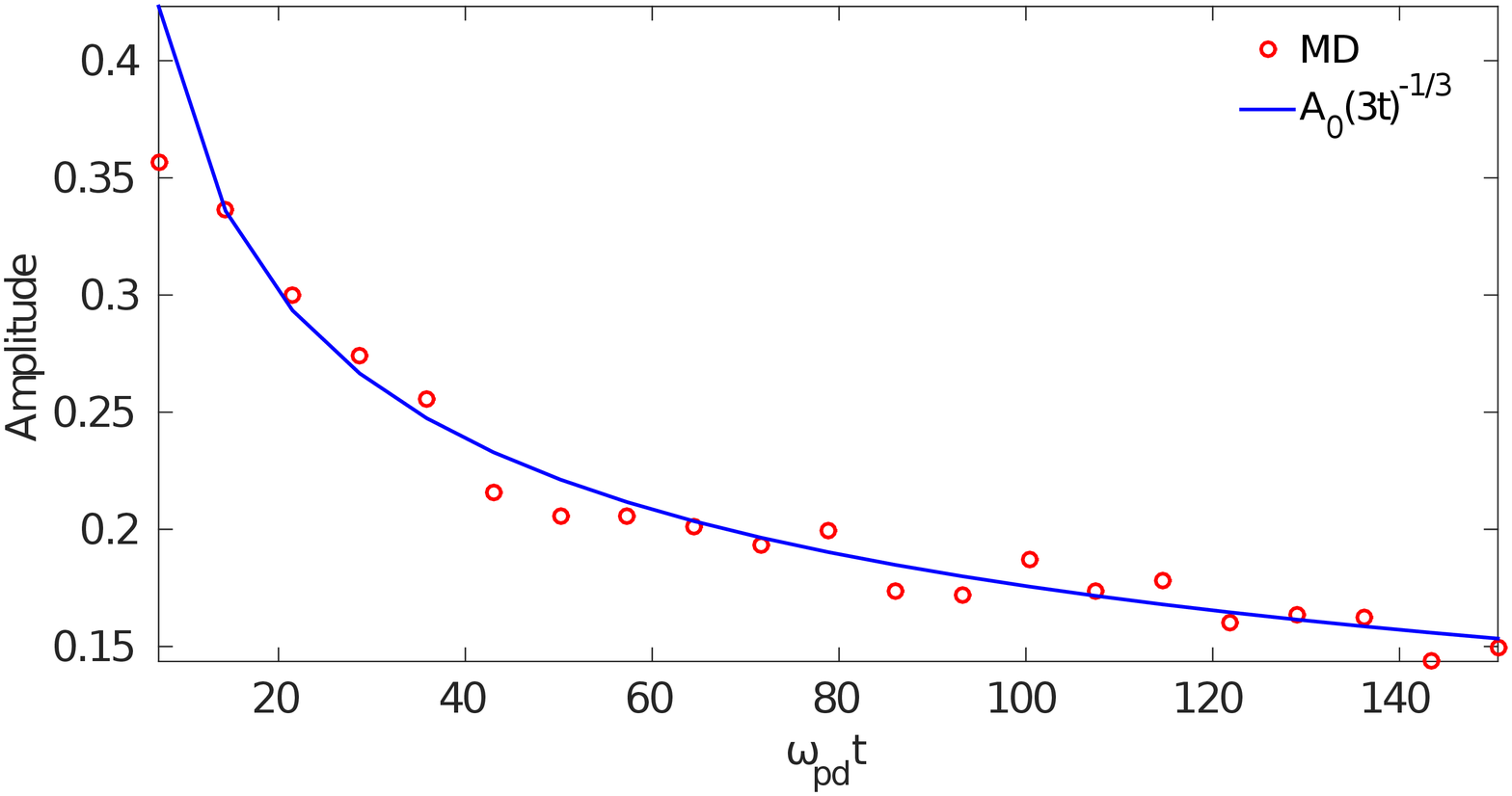}
                   \caption{Comparison of theoretical and simulation results of 
amplitude ($\frac{\delta n}{n_{d0}}$) damping for rarefactive dispersive wave.}
                   \label{damp_wave}	                        
        \end{figure}

We have applied an electric field perturbation along $-\hat{y}$ direction in a narrow rectangular region (to mimic a wire), mathematically $-E\delta(t-t_0)\hat{y}$ at time $t_0$, where $\delta$ is Dirac's delta function. This electric field results in an electrostatic force $F_E = QE\hat{y}$ on the dust particles. The direction of force is along $+\hat{y}$ as we choose the dust charge to be $-Q$ negative. The evolution shows an excitation of a solitary wave propagating in $+\hat{y}$ direction and a damped dispersive wave in the $-\hat{y}$ direction. The time evolution of density ($n_d$) is shown in Fig. {\ref{dispersive}}. These observations are consistent with the property of the solitons permitted by the KdV equation which is given by the equation: 
\begin{equation}
{\frac{\partial n_d}{\partial t} + Cn_d\frac{\partial n_d}{\partial y} + 
D\frac{\partial^3 n_d}{\partial^3 y} = 0}
\end{equation}
Where C and D depends upon density, temperature and mass of particles in the medium\cite{Pintu, Sharma}. 
Second and third term in equation (2) gives the nonlinearity and dispersion in the medium respectively. H. Segur\cite{Segur} and P. G. Drazin\cite{Drazin} have shown that in addition to soliton solutions which is obtained from the balance of nonlinearity and dispersion, negative amplitude dispersive waves solutions propagating in opposite direction with a slower velocity are also permitted. The dispersive waves is the solution of initial-value problem for linearised KdV equation:
\begin{equation}
{\frac{\partial n_d}{\partial t} + D\frac{\partial^3 n_d}{\partial^3 y} = 0}
\nonumber
\end{equation}
The amplitude of such dispersive wave have been shown to decays with time as $A_0\times(3t)^{-\frac{1}{3}}$. Where $A_0$ is the initial amplitude.
Comparison of decaying amplitude of the dispersive wave observed numerically has been provided with the analytic expressions of $A_0\times(3t)^{-\frac{1}{3}}$ in Fig. {\ref{damp_wave}}. It can be seen that there is a close agreement between the two plots.

\begin{figure}[!hbt]   
 \includegraphics[height = 7.0cm,width = 8.0cm]{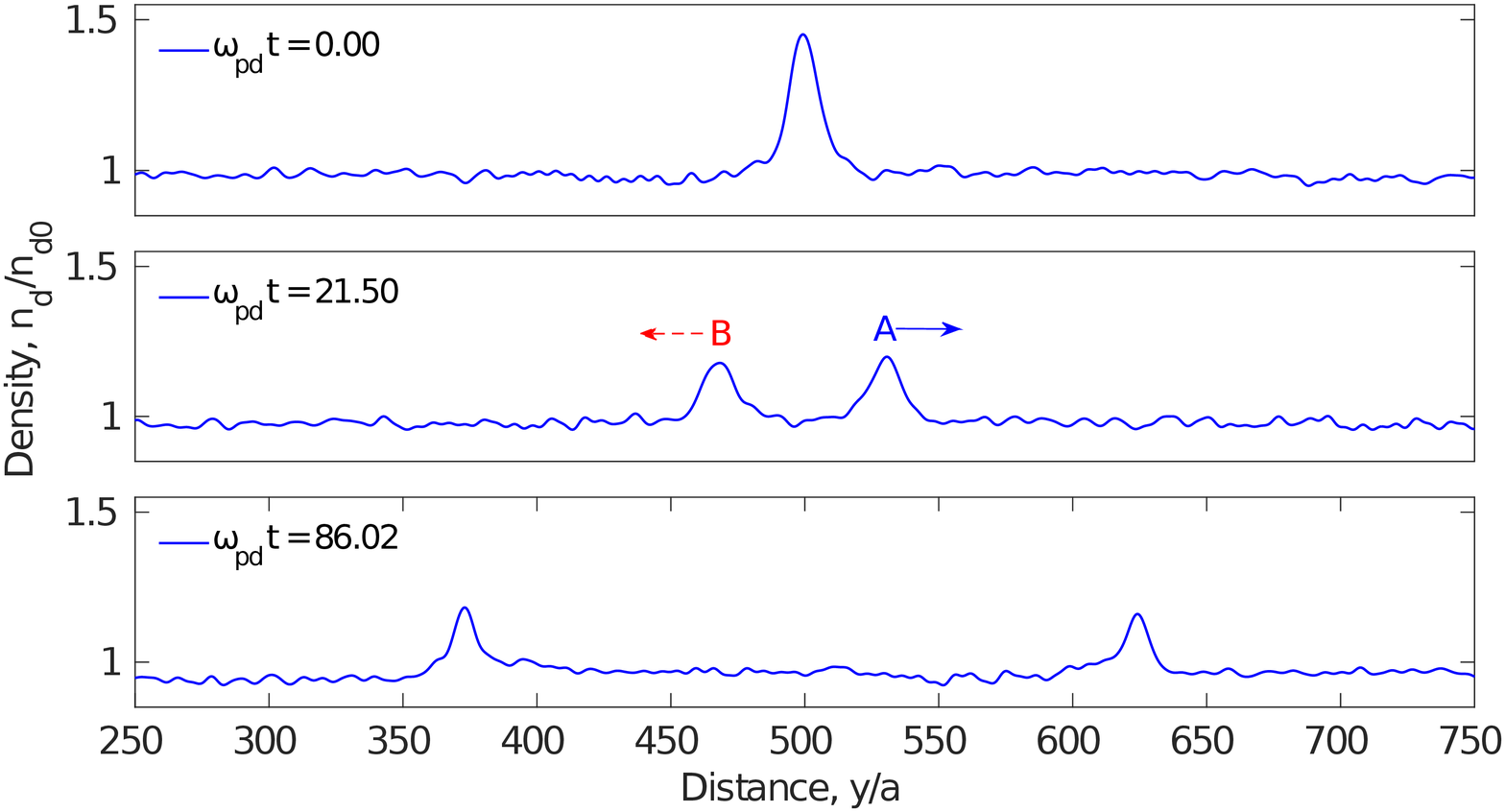}
                   \caption{Time evolution of a Gaussian form of density 
perturbation in the medium. The Gaussian pulse splits in two A ($+\hat{y}$) and B ($-\hat{y}$) oppositely
propagating symmetric pulses due to the left and right symmetry in the medium
                   \cite{sanatsoliton}.}
                  \label{birdsall_gauss}	                         
        \end{figure}

The above observations of propagating solitons in one direction and dispersive wave in other are in contrast to earlier studies carried out by Tiwari et al.\cite{sanatsoliton}, where two oppositely propagating solitonic structures were observed when an arbitrary initial Gaussian density perturbation was evolved. This has been reproduced here by us in Fig. {\ref{birdsall_gauss}}. In the case of an initial Gaussian perturbation in density, there is no way to distinguish between the forward and reverse directions. On the other hand when one considers the response of dust particles with specified charge to an applied electric field the left and right directional symmetry gets broken up.  

Another well known property of KdV soliton is that the parameter $AL^2$ is constant \cite{Pintu, Aossey}. Where $A$ is amplitude and $L$ is full-width at half the maximum amplitude (FWHM) of soliton structure. In Table - I we list some parameters associated with the numerically observed solitonic structures. This include the normalized Electric field amplitude, the Mach number, the normalized value of the soliton width $\frac{L}{a}$, the normalized density amplitude $A = \frac{\delta n}{n_{d0}}$ and $AL^2$ in various columns. The Mach number is the ratio of soliton velocity to the dust acoustic speed, i.e., $M = \frac{v}{C_s}$. The dust acoustic wave speed ($C_s$) of medium at $\Gamma= 100$ and $\kappa = 0.5$ is equal to $1.94\times 10^{-2}$ $(\frac{m}{s})$. The soliton velocity is calculated from the slope of the plot of the soliton trajectory with respect to time. 
\begin{table}[!hbt]
\caption{Soliton Parameter with varying amplitude of perturbation ($E$). Parameters are taken at the time $55.91 \omega_{pd}^{-1}$.}
\begin{center}
 \begin{tabular}{|c c c c c|}
 \hline
 $\frac{E}{E_0}$ &\hspace{0.7cm} $M$ &\hspace{0.7cm} $\frac{L}{a}$ &\hspace{0.7cm} $A$ $(\frac{\delta n}{n_{d0}})$ &\hspace{0.7cm} $AL^2$ \\ [1.5ex] 
 \hline\hline
  20.32 &\hspace{0.7cm} $1.15$ &\hspace{0.7cm} 9.5 &\hspace{0.7cm}  0.307 &\hspace{0.7cm}       27.70\\ 
 \hline
 22.86 &\hspace{0.7cm} $1.16$ &\hspace{0.7cm} 9.0 &\hspace{0.7cm} 0.324 &\hspace{0.7cm} 26.24 \\
 \hline
 25.40 &\hspace{0.7cm} $1.18$ &\hspace{0.7cm} 8.5 &\hspace{0.7cm} 0.346 &\hspace{0.7cm} 24.99 \\
 \hline
 28.96 &\hspace{0.7cm} $1.20$ &\hspace{0.7cm} 8.5 &\hspace{0.7cm} 0.365 &\hspace{0.7cm} 26.37 \\
 \hline
 30.49 &\hspace{0.7cm} $1.22$ &\hspace{0.7cm} 8.1 &\hspace{0.7cm} 0.393 &\hspace{0.7cm} 25.78 \\ [1.5ex] 
 \hline
\end{tabular}
\end{center}
\label{sol_para_table}
\end{table} 
From table - {\ref{sol_para_table}}, we find that with increasing amplitude ($A$) the width ($L$) of solitary wave decreases as expected. From the table, it is also clear that soliton parameter $AL^2$ remains fairly constant for solitons with different Mach numbers ($M$). This can be understood from the fact that even though the percentage variation in the data between the minimum and maximum value of $L^2$ is about $27 \%$, in $A$ about $21\%$; the variation in $AL^2$ is limited to $7\%$ only. This can be attributed to be well within the inaccuracy in estimation. 

\section{Interaction between solitons} 
\label{solcol}
We also report on the collisional interaction characteristics of the numerically evolved structures which shows clear solitonic behaviour. 
\subsection{Head-on collision of same amplitude solitons:} 
\begin{figure}[!hbt]   
 \includegraphics[height = 7.0cm,width = 8.0cm]{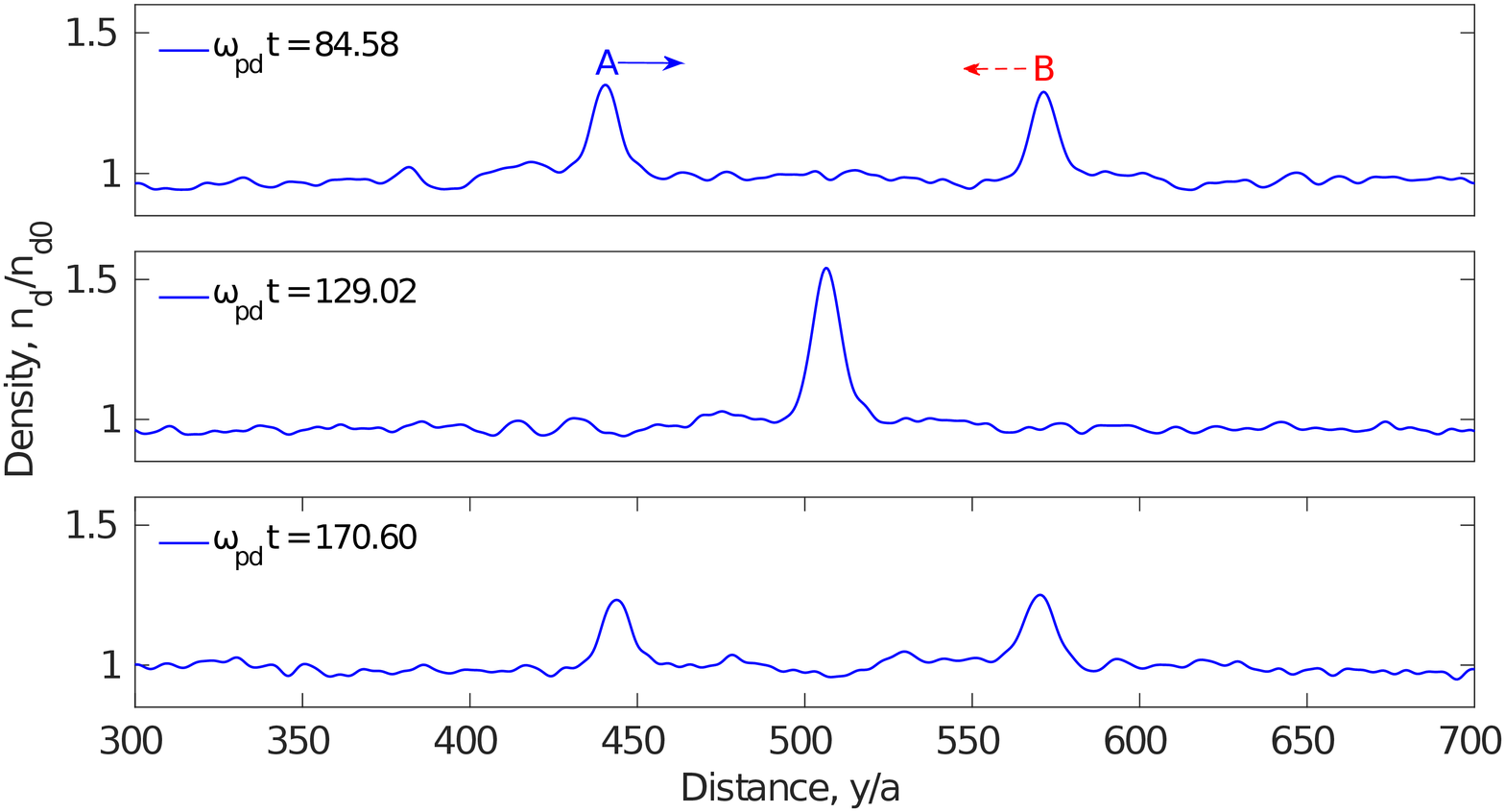}
                   \caption{
                    Collision of two same amplitude counter 
                    propagating solitons. Oppositely moving solitons (A and 
                    B) were excited with same amplitude electric field ($E = 
                    25.40$) in $+\hat{y}$ and $-\hat{y}$ directions respectively.}
                   \label{head_coll_2500}	                       
        \end{figure}

\begin{figure}[!hbt]
 \includegraphics[height = 7.0cm,width = 8.0cm]{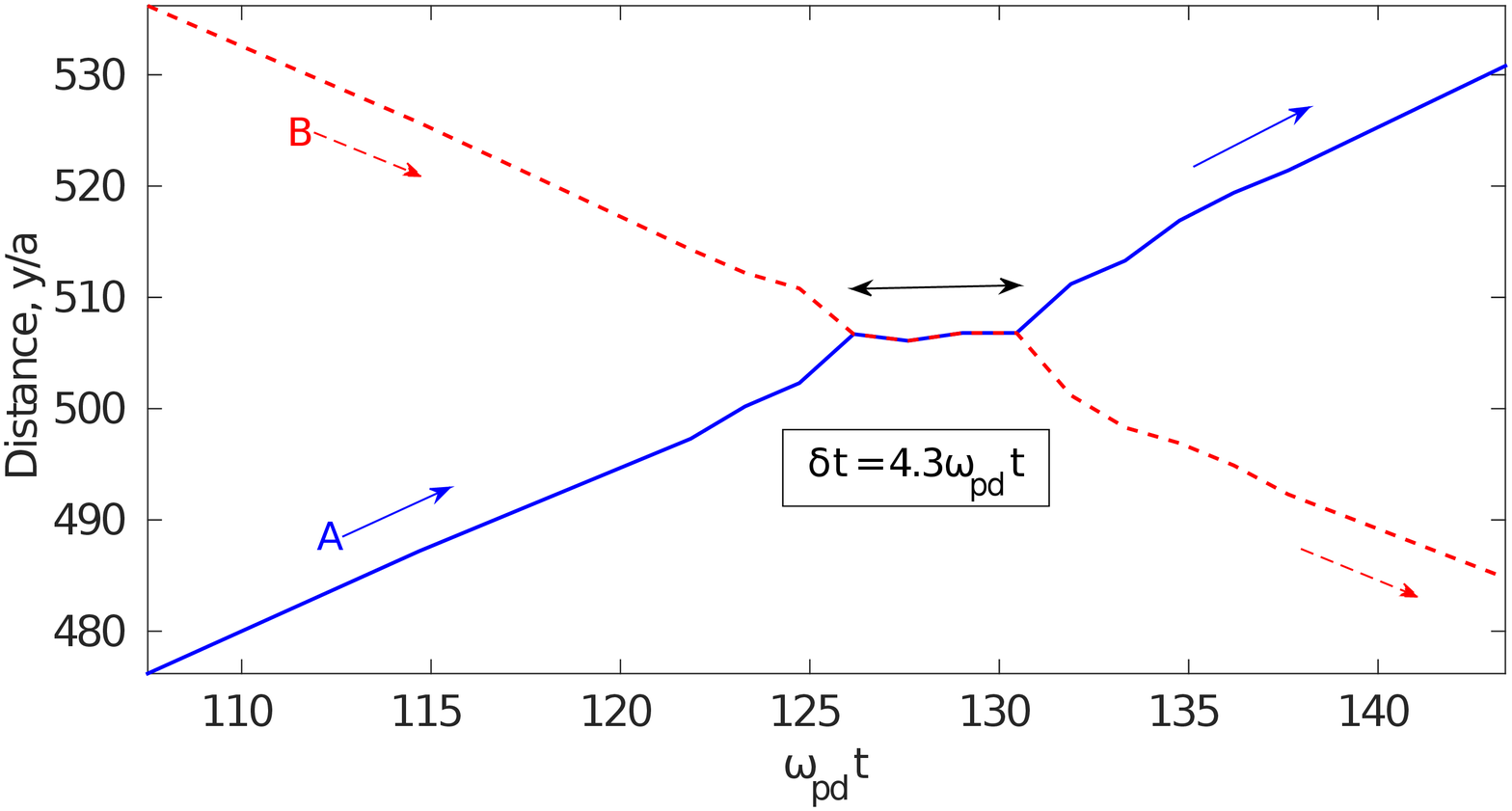}
                   \caption{Phase shift during same amplitude soliton collision. An initial electric field to excite them is $E = 25.40$.}
                   \label{phase_shift_2500}	                         
        \end{figure}

\begin{figure}[!hbt]   
 \includegraphics[height = 7.0cm,width = 8.0cm]{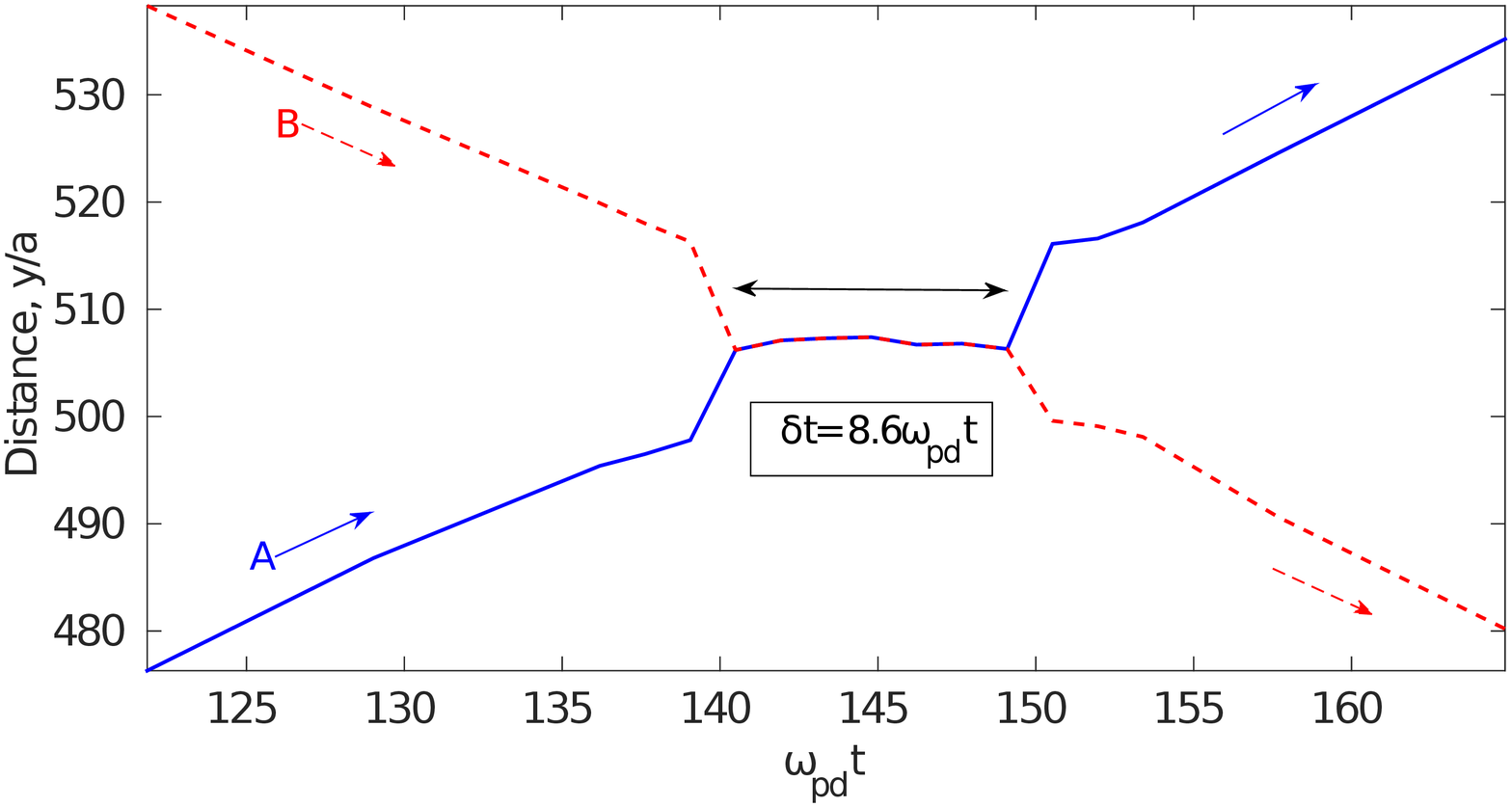}
                   \caption{Phase shift during same amplitude soliton collision. An initial electric field to excite them is $E = 12.70$.}
                   \label{phase_shift_1250}	                       
        \end{figure}

By applying suitable electric field perturbations at different locations we create two counter propagating solitons of same amplitude. Time evolution of these structures are shown in Fig. {\ref{head_coll_2500}}. The structures collide and cross each other with no change in their shape and size. We also observe that during the time they overlap while colliding the resultant amplitude ($0.589$) of solitary wave is less than the sum of the individual soliton amplitudes ($0.318 + 0.318 = 0.636$).  
The trajectories of the two solitons with initial amplitudes of perturbation $E = 25.40$ and $E = 12.70$ are shown in Fig. {\ref{phase_shift_2500}} and Fig. {\ref{phase_shift_1250}} respectively. Since the solitons are of equal amplitude the structures remain static for some time when they overlap. Time difference ($\delta t$) between the two points (intersection of incoming and outgoing trajectories) is termed as phase shift and it is about $4.3\omega_{pd}^{-1}$ and $8.6\omega_{pd}^{-1}$ for the two cases as shown in Fig. {\ref{phase_shift_2500}} and Fig. {\ref{phase_shift_1250}} respectively. The phase shift clearly decreases with increase in the amplitude of the solitons. This particular result is in contrast to the experimental findings of Sharma et al.\cite{Sharma}. The reason for this difference is not clear at the moment. However, intuitively one would expect the collision between higher amplitude solitons to have smaller phase shifts as they move with greater speeds. Conclusion similar to ours on phase shift has been theoretically inferred in some previous studies \cite{Aossey, Verheest}.  


\subsection {Head-on collision of different amplitude solitons:} 
\begin{figure}[!hbt] 
 \includegraphics[height = 7.0cm,width = 8.0cm]{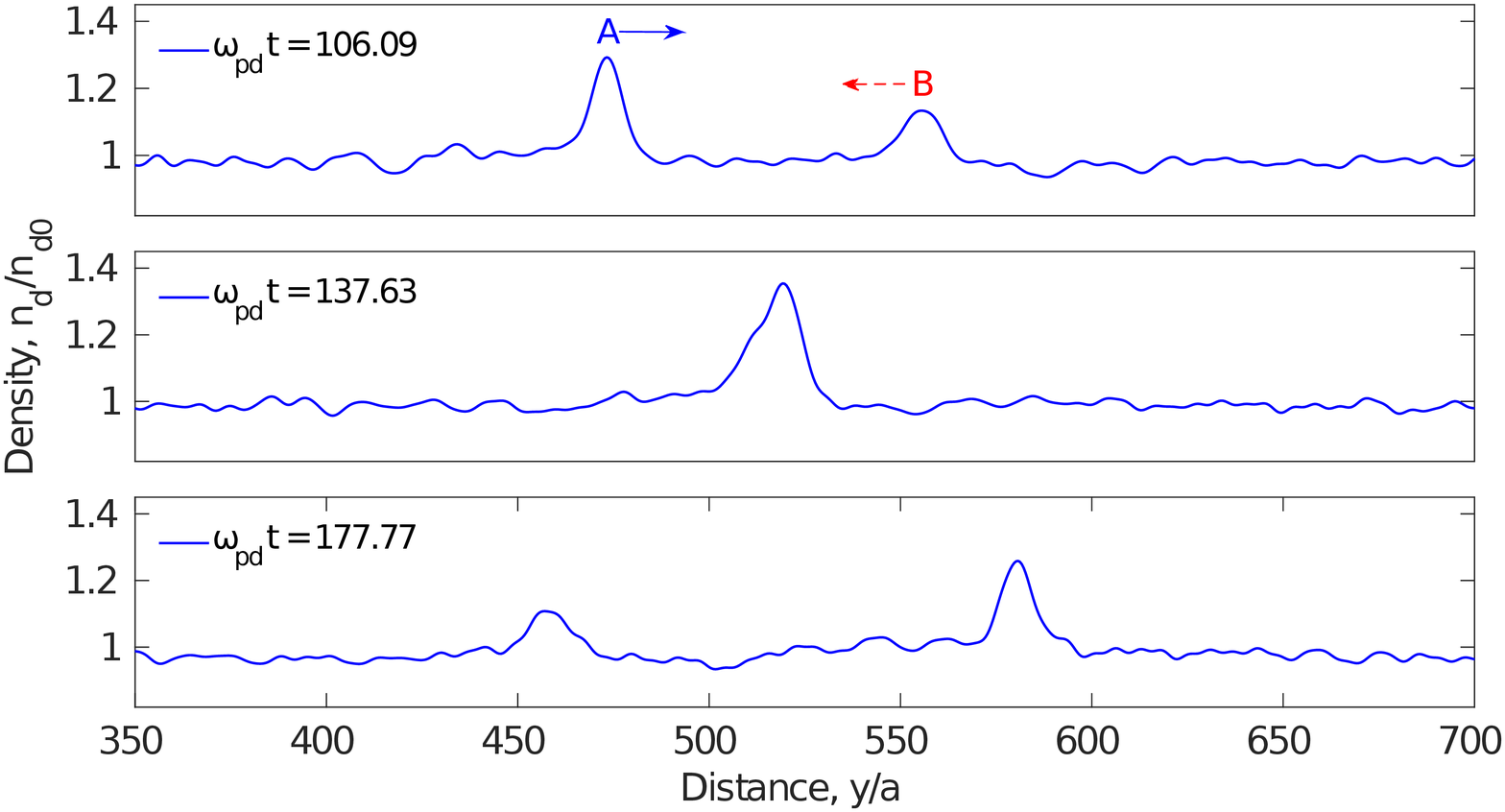}
                   \caption{Head-on collision of different amplitude solitons A ($E = 25.40$) and B ($E = 15.24 $) moving in opposite direction.}
                    \label{head_coll_2500_1500}                        
        \end{figure}

\begin{figure}[!hbt] 
 \includegraphics[height = 7.0cm,width = 8.0cm]{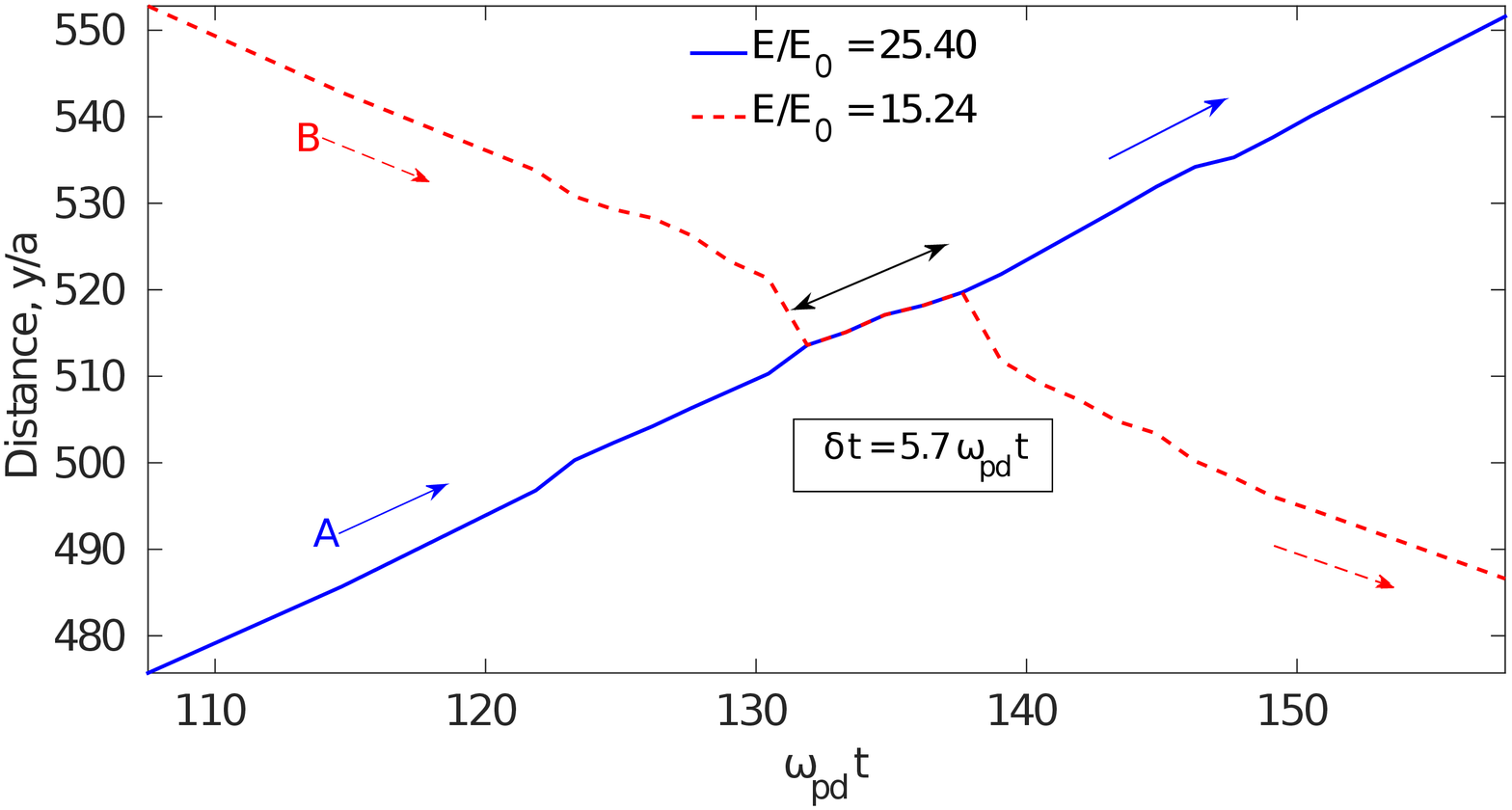}
                   \caption{Phase shift for head-on collision of different amplitude
                   solitons A ($E = 25.40$) and B ($E = 15.24$).}
                   \label{phase_shift_2500_1500} 	                        
 
        \end{figure}

We have also considered the case of head on collision amidst two counter propagating solitonic structures with unequal amplitude. Again the two structures emerge unchanged 
after suffering collision as shown in Fig. {\ref{head_coll_2500_1500}}. From the plot of Fig. {\ref{phase_shift_2500_1500}}, which shows the trajectories of the two solitons, it can 
be observed that after the collision the low amplitude soliton in this case gets dragged in the direction opposite to its own propagation, by the high amplitude structure for a while.  
This is in confirmation with the analytical results obtained by Surabhi et al.\cite{Jaiswal}. This time phase shift is about $5.7 \omega_{pd}^{-1}$ which is shown in Fig. {\ref{phase_shift_2500_1500}}. Thereafter the two structures get separated and move in their respective directions. In this case too the resultant amplitude ($0.436$) of the structure during collision is less than the sum of the individual amplitudes ($0.308 + 0.185 = 0.493$) of the solitons.  
 \\

\subsection{Overtaking collision amidst different amplitude solitons:}
\begin{figure}[!hbt] 
 \includegraphics[height = 7.0cm,width = 8.0cm]{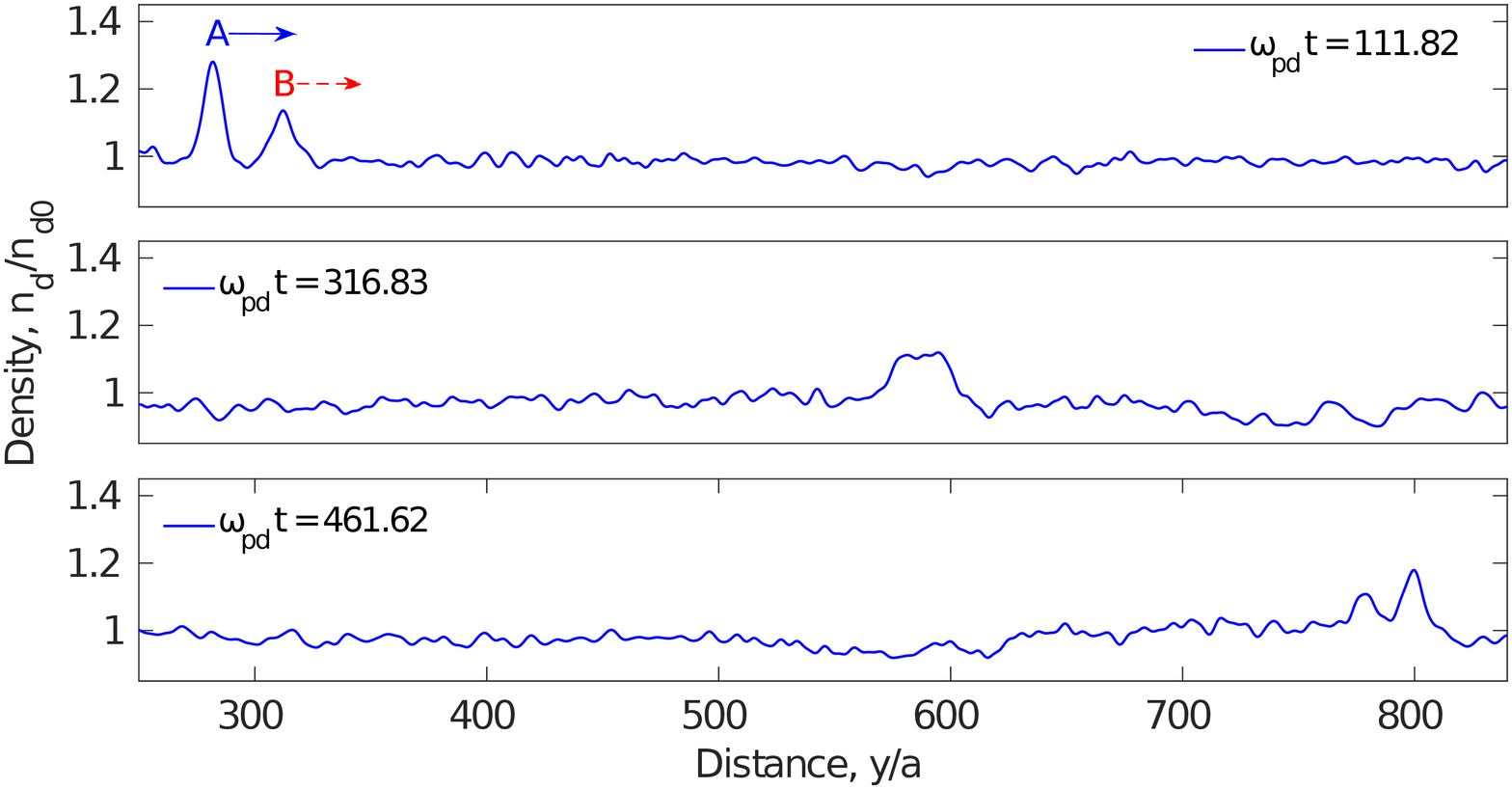}
                  \caption{Density evolution for overtaking collision of A ($E = 
25.40$) and B ($E = 15.24$) amplitude solitons.}
        \label{over_coll_2500_1500}
        \end{figure}

In this case we excite two solitons propagating in the same directions. The smaller amplitude soliton with slower phase velocity is placed ahead of the high amplitude soliton which is moving faster. After some time the faster soliton catches up with the slower soliton ahead of it and collides with it. This has been shown in Fig. {\ref{over_coll_2500_1500}}. We have found that phase shift in the overtaking collision is larger than head-on collision. 

\section{Excitation of Multi-solitons and effect of the coupling parameter:}
\label{multisol}
\begin{figure}[!hbt]
        \centering   
 \includegraphics[height = 7.0cm,width = 8.0cm]{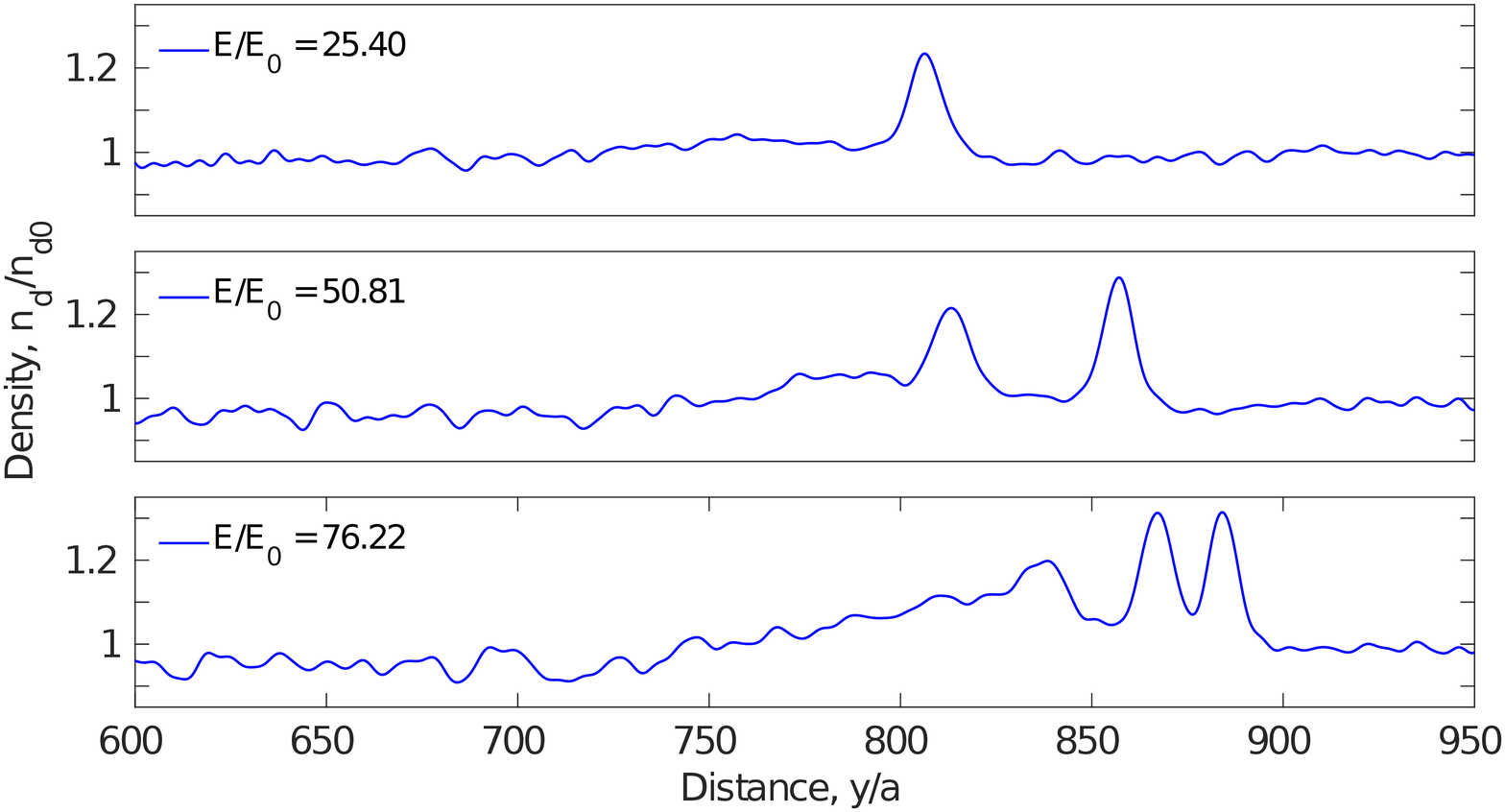}
                   \caption{Formation of multi-soliton due to the increase in electric
                    field strength ($E$). Density of medium for all $E$ is taken at time 
                    $263.78 \omega_{pd}^{-1}$. In all three cases perturbation width ($d$)
                     is 10.}
                  \label{diff_E}	                        
 
        \end{figure}

\begin{figure}[!hbt]
        \centering   
 \includegraphics[height = 7.0cm,width = 8.0cm]{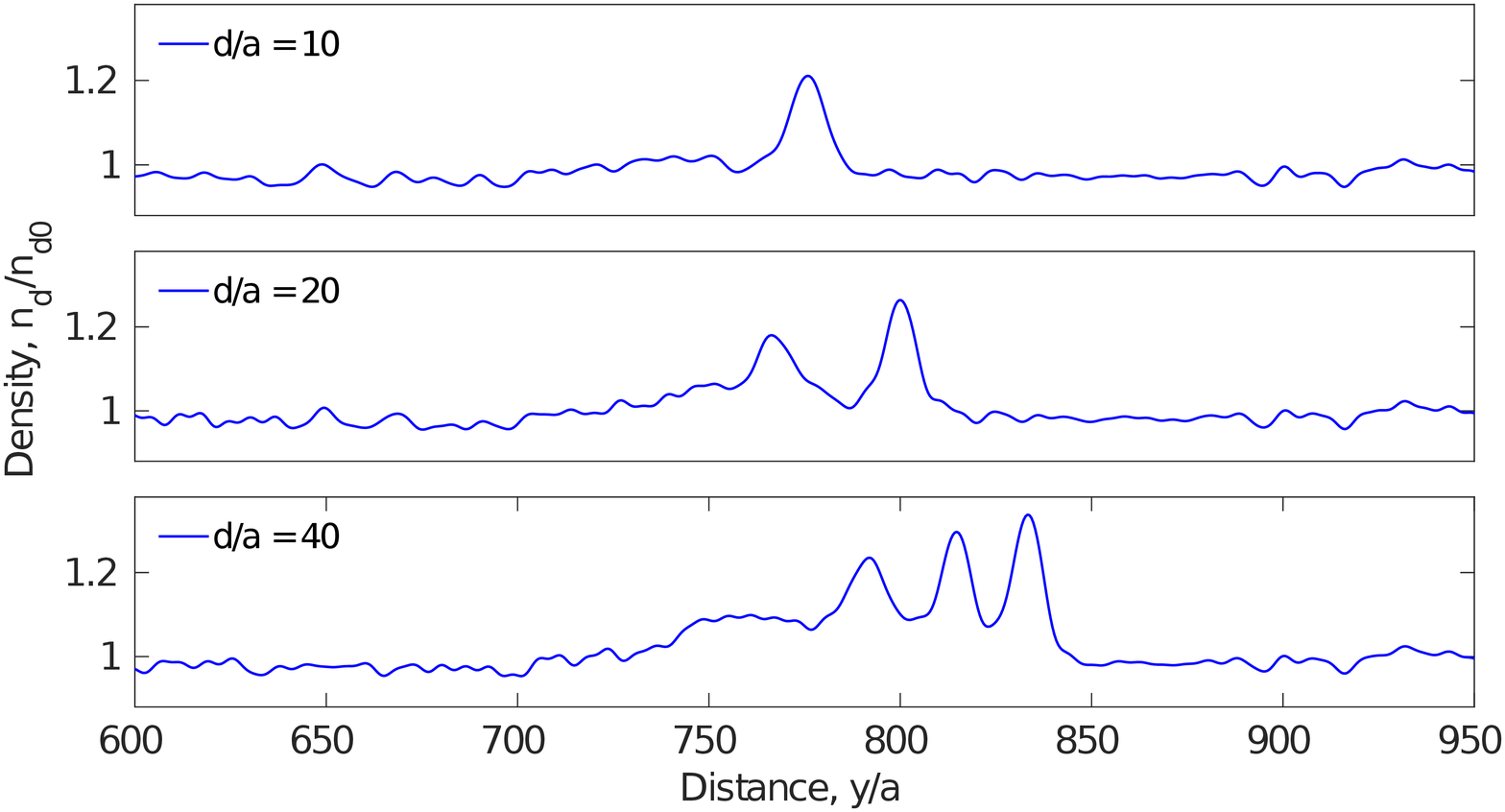}
                   \caption{Formation of multi-soliton due to the increase in
                    perturbation width ($d$). Density of medium for all $d$ is taken at time 
                    $242.28 \omega_{pd}^{-1}$. In all three cases perturbation 
                    strength of electric field ($E$) is $25.40$.}
                   \label{diff_d}	                        
 
        \end{figure}

\begin{figure}[!hbt]
        \centering   
 \includegraphics[height = 7.0cm,width = 8.0cm]{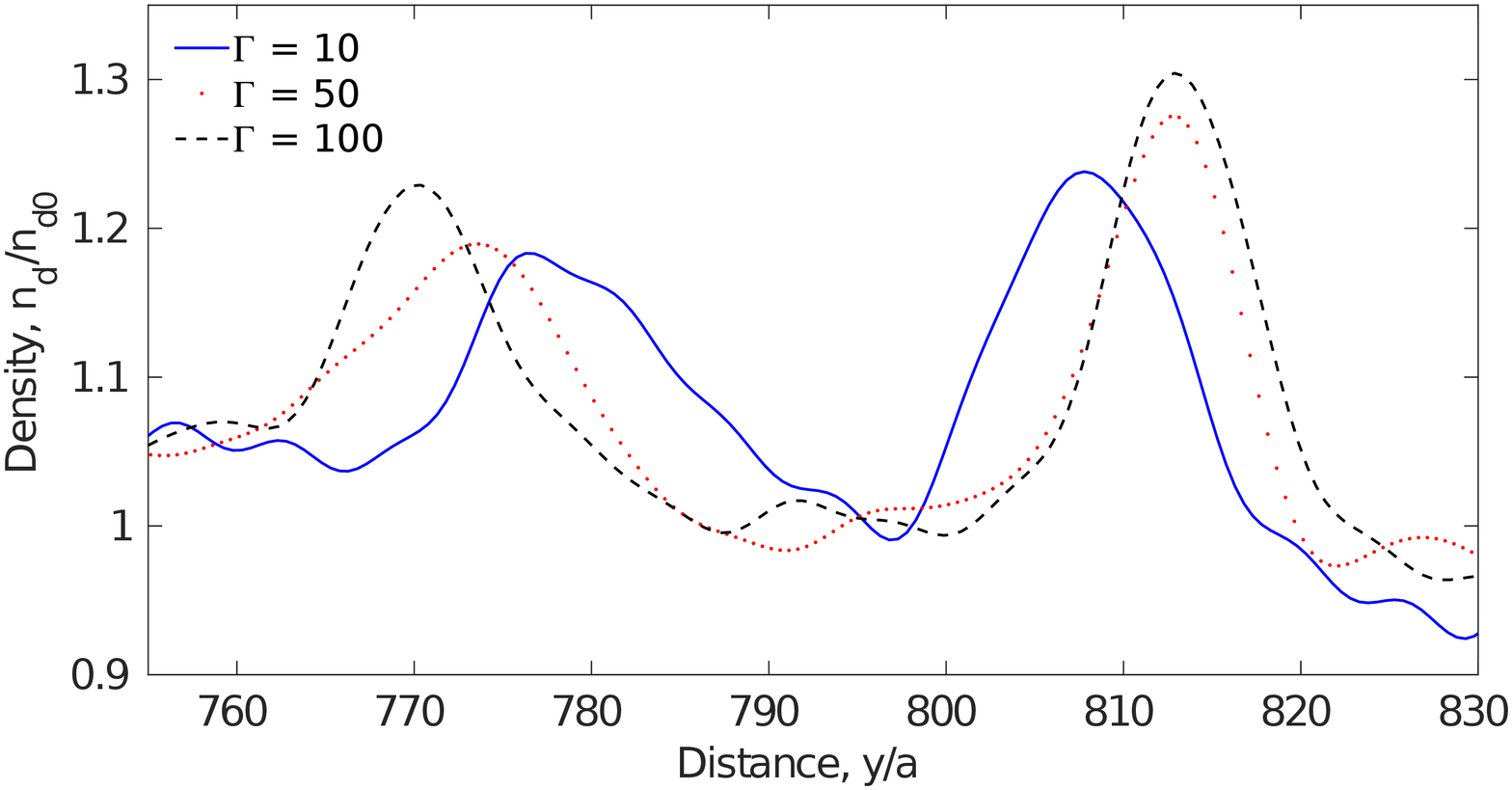}
                   \caption{Density of medium for three $\Gamma = $ 10, 50 and 100 is taken at time 
                    $235.11 \omega_{pd}^{-1}$. For all three cases magnitude and width of electric field perturbation is $50.81$ and $10$ respectively.}
                   \label{diff_gm}	                        
 
        \end{figure}

When we increase either the amplitude ($E$) or the spatial width ($d$) of the electric field impulse, the solitary pulse excited in the forward $+\hat{y}$ direction breaks up into more than one solitons. In Fig. {\ref{diff_E}} it is shown that for a fixed value of $d=10$, as the amplitude of electric field perturbation is increased, multiple solitons appear. Similarly when the electric field amplitude is fixed and the width $d$ is increased, multiple solitons get formed as shown in Fig. {\ref{diff_d}}. Each of these multiple structures propagate along the same direction. They arrange themselves in the order of decreasing amplitude $A$ (increasing width $L$). Interestingly the crests of each of the structures lie close to a straight line. These multiple solitons are termed as multi-solitons. In one of the recent experiments done by Boruah et al.\citep{Boruah} of dusty plasmas the multi-soliton formation has been clearly demonstrated by increasing the electric field impulse. The MD simulations with Yukawa interaction thus seems to be a good depiction of the properties of the dusty plasma medium. It should be noted that the formation of multiple solitons had been theoretically predicted by Zabusky et al. in the context of electron-ion plasma\cite{Zabusky}.\\
We have also investigated the role of coupling parameter on the formation of these soliton structures. We observe that with increasing coupling parameter $\Gamma$ of the dust medium, amplitude (magnitude) of each soliton increases and consequently the width decreases as shown in Fig. {\ref{diff_gm}}.

\section{Summary} 
\label{result}
We have carried out the MD simulations for dusty plasma medium, treating the medium as collection of dust particles interacting with Yukawa interaction. We study the response of the dust medium to an imposed electric field impulse and provide clear evidence of the formation of KdV solitons. These evidences are in terms of following features: (a) a creation of positive density pulse propagating in one direction along with a train of negative perturbed density oscillations in the opposite direction (b) relative constancy of $AL^2$ (here $A$ is the amplitude and $L$ is the full width at half maxima of the structure) (c) the structures are shown to be remain intact after undergoing collisional interaction amidst them. Interestingly, the results (a) and (b) are supported by an experimental observation of dusty plasma made by Sheridan et al.\cite{Sheridan}. 

We have also demonstrated that by increasing the strength of electric field impulse the amplitude of the solitonic structure increases and after a point it starts to splits in the form of multiple solitons. This is in agreement with recent experiments which have reported the formation of multiple solitons\citep{Boruah}. This suggests that the depiction of the dusty plasma medium in terms of a simple model of a collection of dust particle interaction via Yukawa potential is fairly good. Another observation made in the present study is related to studying the role of coupling parameter on the formation of solitonic structures. We have shown that by increasing the coupling parameter of the medium the amplitude of the solitonic structures increases while its width decreases. 
 
Furthermore, we have observed that the phase shifts in the collisional interaction seems to decrease with the increasing amplitude of the colliding solitonic structures. In one recent experimental observations\cite{Sharma} as well as in some other literature\cite{Jaiswal} contrary to this has been reported. We feel that our observations appears consistent with intuition, as one would expect the interaction time between two rapidly moving solitons (which have higher amplitude) to be smaller 
compared to slowly moving low amplitude solitons. We, therefore, feel that a relook of this issue 
in experiments as well as theoretical analysis is necessary. 

\newpage
\bibliographystyle{unsrt}

%
\end{document}